\documentclass{article}
\usepackage{graphicx}

\newcommand{\CMSSMI}{CMSSM-seesaw~I}

\newcommand{\fh}{{\tt FeynHiggs}}

\newcommand{\MW}{M_W}

\newcommand{\MH}{M_H}

\newcommand{\tsf}{\theta\kern-.20em_{\tilde{f}}}
\newcommand{\tsfp}{\theta\kern-.20em_{\tilde{f}\prime}}
\newcommand{\tsq}{\theta\kern-.15em_{\tilde{q}}}
\newcommand{\sweff}{\sin^2\theta_{\mathrm{eff}}}

\newcommand{\VL}{\left( \begin{array}{c}}
\newcommand{\VR}{\end{array} \right)}
\newcommand{\ML}{\left( \begin{array}{cc}}
\newcommand{\MLd}{\left( \begin{array}{ccc}}
\newcommand{\MLv}{\left( \begin{array}{cccc}}
\newcommand{\MR}{\end{array} \right)}

\newcommand{\tb}{\tan \beta}

\newcommand{\gev}{\,\, \mathrm{GeV}}
\newcommand{\mev}{\,\, \mathrm{MeV}}

\newcommand{\BC}{\begin{center}}
\newcommand{\EC}{\end{center}}
\newcommand{\BE}{\begin{equation}}
\newcommand{\EE}{\end{equation}}
\newcommand{\BEA}{\begin{eqnarray}}
\newcommand{\EEA}{\end{eqnarray}}

\newcommand{\id}{{\rm 1\kern-.12em
\rule{0.3pt}{1.5ex}\raisebox{0.0ex}{\rule{0.1em}{0.3pt}}}}

\def\ga{\gamma}

\def\de{\delta}

\newcommand{\deFABij}{\de^{FAB}_{ij}}

\def\De{\Delta}

\newcommand{\bsg}{\ensuremath{\br(B \to X_s \ga)}}

\newcommand{\bmm}{\ensuremath{\br(B_s \to \mu^+\mu^-)}}
\newcommand{\dmbs}{\ensuremath{\De M_{B_s}}}

\newcommand{\DMH}{\ensuremath{\De \MH^{\rm MFV}}}

\newcommand{\DMW}{\ensuremath{\De\MW^{\rm MFV}}}
\newcommand{\Dsweff}{\ensuremath{\De\sweff^{\rm MFV}}}

\newcommand{\br}{{\rm BR}}

\begin{document}
\title{ 
Effects of Sfermion Mixing induced by RGE in the CMSSM
}
\author{ Mario E. G\'omez       \\
{\em Department of Applied Physics, University of Huelva, 21071 Huelva, Spain } \\
Sven Heinemeyer  \\
{\em Instituto de F\'isica de Cantabria (CSIC-UC),  39005 Santander, Spain}\\
Muhammad Rehman \\
{\em Instituto de F\'isica de Cantabria (CSIC-UC),  39005 Santander, Spain}
}
\maketitle
\baselineskip=11.6pt
\begin{abstract}
Even within the Constrained Minimal Supersymmetric Standard Model (CMSSM) it is possible to induce sfermion flavor mixing through the Renormalization Group Equations (RGE) when the full  structure of the Yukawa couplings is considered. We analyse the impact of including those effects  on the accurate computation of  $B$-physics observables, electroweak precision observables (EWPO) and the Higgs boson
mass predictions. 
\end{abstract}
\baselineskip=14pt
\section{Introduction}
Supersymmetric (SUSY) extensions of the Standard
Model (SM) \cite{mssm} come with many promises to become the next step in the search for new physics. They offer a solution to the hierarchy  problem, a candiate to Dark Matter and many new particles at the range of the energy of the LHC. Now, with the data of the first run of the LHC we have new bounds for the SUSY observables which constraint  the SUSY parameters. We study the impact of these bounds to the SUSY contribution to some of the SM well measured  observables. In particular, we include on our analysis  flavor violating  (FV) contributions  arising from the new  SUSY particles. 

We work in the framework of the  Minimal Supersymmetric extension of the SM (MSSM), with the additional assumption that SUSY is broken by universal soft terms at the grand unification scale (GUT). In this framework, called constrained MSSM (CMSSM), there is FV only in the squark sector. This arises due to the presence of the Yukawa couplings in the RGE's, such that  the FV is only due the CKM matrix.  Hence,  this is the Minimal Flavor Violation (MFV) scenario \cite{MFV1}. However, this is not enough to explain the experimental evidence for neutrino flavor oscillations. In order to account for those, we must enlarge the CMSSM.  This can  achieved by augmenting the CMSSM with  a ''see-saw'' mechanism od type I. The resulting model, called "CMSSM-seesaw I'' predicts also FV in the lepton sector (LFV).    

We present in the next sections some results on FV predictions on the CMSSM and their contribution to the evaluation of electroweak precision observables (EWPO), in particular $\MW$ and the
effective weak leptonic mixing angle, $\sweff$. The effects on other observables like  $B$~physics
observables (BPO), in particular \bsg, \bmm\ and \dmbs,  as well as the masses of
the neutral and charged Higgs bosons in the MSSM where found to be small.  We refer the reader to ref.~\cite {us1,us2} for furher details of our computation and a complete list of references.

\begin{figure}[ht!]
\begin{center}
  {\includegraphics[scale=0.44]{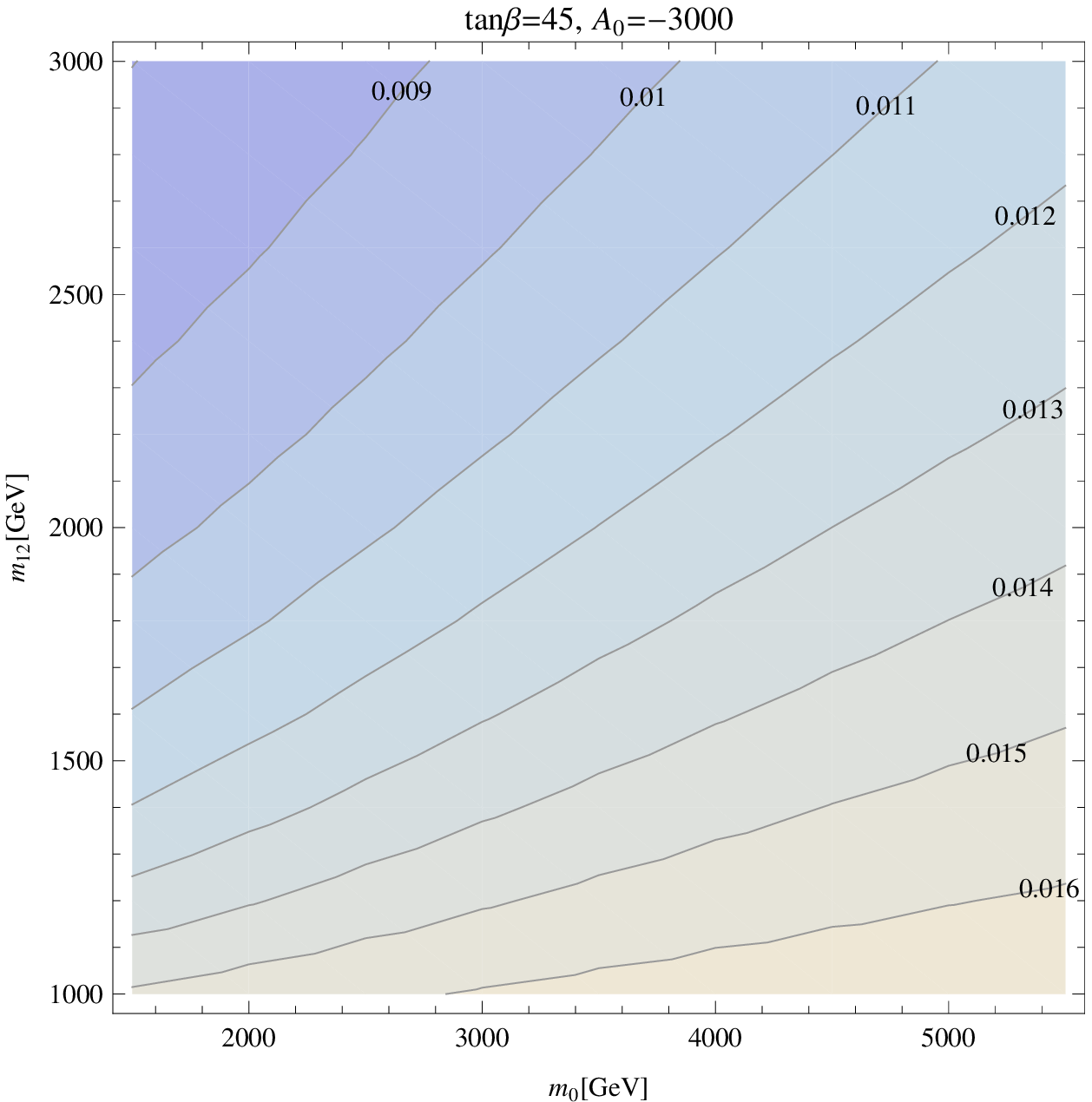}}\hspace{0.5cm}
        {\includegraphics[scale=0.44]{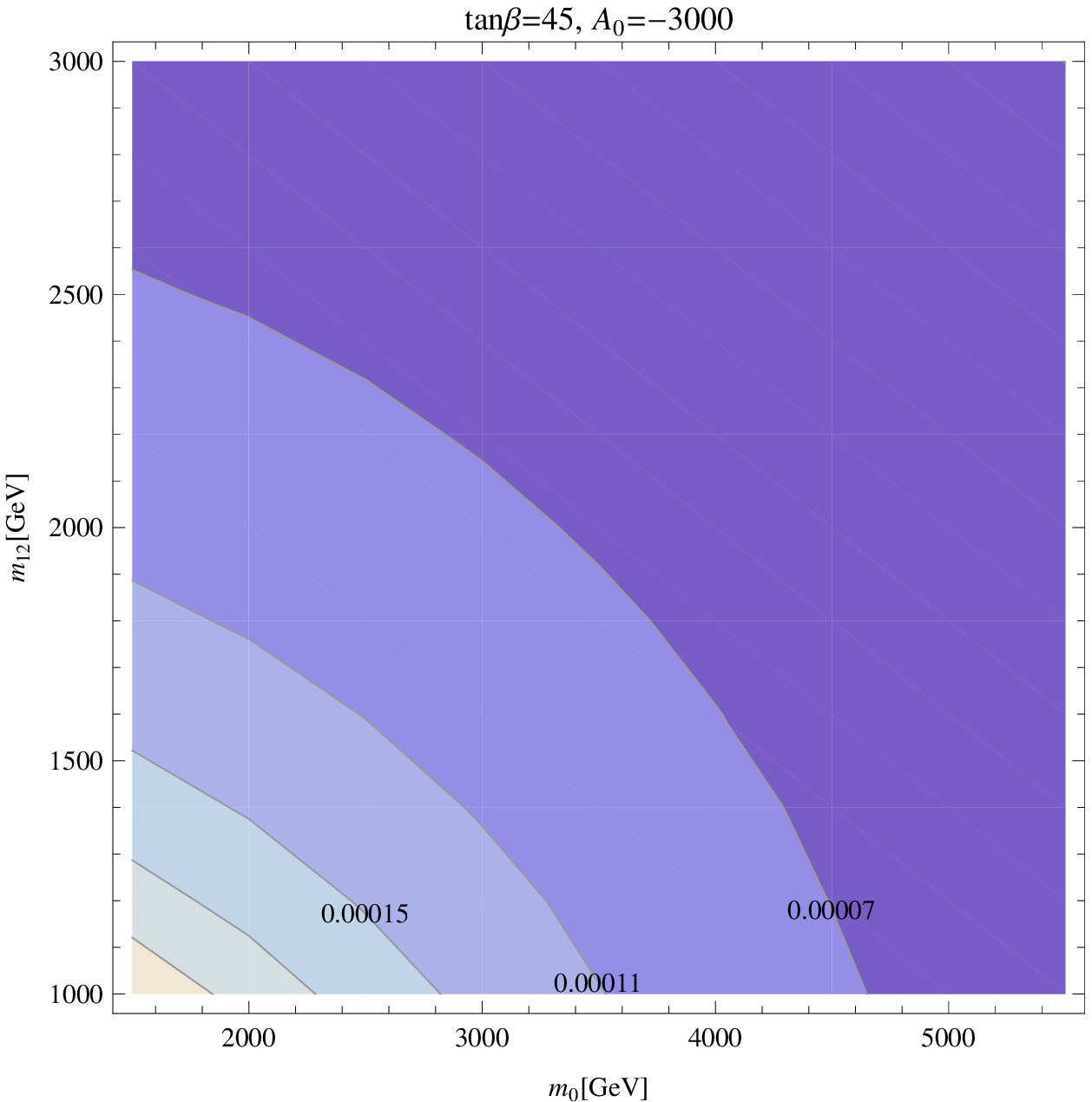}}
\end{center}
\caption{\it Contours of $\delta^{QLL}_{23}$ (left) and $\delta^{ULR}_{23}$ (right) in the
  $m_0$--$m_{1/2}$ plane for $\tb=45$ and    
$A_0=-3000$~GeV in the CMSSM.}  
\label{fig:DelQLL23}
\end{figure} 

\section{Scalar fermion sector with flavor mixing}
\label{sec:sfermions}

The MSSM is defined by the superpotential:
\begin{equation}
\label{superpotential}
W_{\rm MSSM}=\epsilon_{\alpha \beta}(Y_e^{ij}H_1^{\alpha} E_i^c  L_j^{\beta}
+ Y_{d}^{ij}  H_1^{\alpha} D_i^c  Q_j^{\beta}
+ Y_{u}^{ij}  H_2^{\alpha} U_i^c Q_j^{\beta}
+ \mu  H_1^{\alpha} H_2^{\beta})
\end{equation}
where $L_i$ represents the chiral multiplet of a $SU(2)_L$ doublet
lepton, $E_i^c$ a $SU(2)_L$ singlet charged lepton, $H_1$ and $H_2$ two Higgs doublets with opposite hypercharge.
Similarly $Q$, $U$ and $D$ represent chiral multiplets of quarks of a
$SU(2)_L$ doublet and two singlets with different $U(1)_Y$ charges.
Three generations of leptons and quarks are assumed and thus the
subscripts $i$ and $j$ run over 1 to 3. The symbol $\epsilon_{\alpha
\beta}$ is an anti-symmetric tensor with $\epsilon_{12}=1$.  SUSY is ''softly brooken" by a scalar potential with  bilinear and trilinear combinations of  the superpartners. Within the Constrained MSSM the soft SUSY-breaking parameters are assumed to be universal at the Grand Unification scale
$M_{\rm GUT} \sim 2 \times 10^{16} \gev$. All the scalars are assumed to have the same mass, $m_0$, the trilinear soft tems are proportional to their respective  Yukawa couplings and fermionic partners of the gauge bosons have a common mass $m_{1/2}$. Since the soft terms are universal, at the GUT scale, they are invariant under superfield rotations.  Hence, it is possible  to work in the basis in which the Yukawa couplings are  $Y_{D} = {\rm diag}(y_{d},y_{s},y_{b})$ and $Y_{U} = V_{\rm CKM}^{\dag} {\rm diag}(y_{u},y_{c},y_{t})$ such that FV terms display an explicit dependence on the CKM matrix.  

The SUSY spectra have been generated with the code 
{\tt SPheno 3.2.4}~\cite{Porod:2003um}.  All the SUSY masses and mixings are then given as 
a function of $m_0^2$, $m_{1/2}$, $A_0$, and 
$\tb = v_2/v_1$, the ratio of the two vacuum expectation values (see
below). We require radiative symmetry breaking to fix $|\mu|$ and 
$|B \mu|$ with the tree--level Higgs potential.  The non-diagonal entries in this $6 \times 6$ general matrix for sfermions
can be described in terms of a set of 
dimensionless parameters $\deFABij$ ($F=Q,U,D,L,E; A,B=L,R$; $i,j=1,2,3$, 
$i \neq j$) where  $F$ identifies the sfermion type, $L,R$ refer to the 
``left-'' and ``right-handed'' SUSY partners of the corresponding
fermionic degrees of freedom, and $i,j$
indexes run over the three generations. The soft sfermion mass matrices in terms of the $\deFABij$ are 
\begin{equation}  
m^2_{\tilde U_L}= \left(\begin{array}{ccc}
 m^2_{\tilde Q_{1}} & \de_{12}^{QLL} m_{\tilde Q_{1}}m_{\tilde Q_{2}} & 
 \de_{13}^{QLL} m_{\tilde Q_{1}}m_{\tilde Q_{3}} \\
 \de_{21}^{QLL} m_{\tilde Q_{2}}m_{\tilde Q_{1}} & m^2_{\tilde Q_{2}}  & 
 \de_{23}^{QLL} m_{\tilde Q_{2}}m_{\tilde Q_{3}}\\
 \de_{31}^{QLL} m_{\tilde Q_{3}}m_{\tilde Q_{1}} & 
 \de_{32}^{QLL} m_{\tilde Q_{3}}m_{\tilde Q_{2}}& m^2_{\tilde Q_{3}} 
\end{array}\right)~,
\label{mUL}
\end{equation}
 \noindent
$m^2_{\tilde U_R}$  and $m^2_{\tilde D_R}$ are defined in a similar way, while   $m^2_{\tilde D_L}$ verifies:  $m^2_{\tilde D_L}= V_{\rm CKM}^\dagger \, m^2_{\tilde U_L} \, V_{\rm CKM}$. The trilinear terms can be written as:
 
\noindent 
\begin{equation}
v_2 {\cal A}^u  =\left(\begin{array}{ccc}
 m_u A_u & \de_{12}^{ULR} m_{\tilde Q_{1}}m_{\tilde U_{2}} & 
 \de_{13}^{ULR} m_{\tilde Q_{1}}m_{\tilde U_{3}}\\
 \de_{{21}}^{ULR}  m_{\tilde Q_{2}}m_{\tilde U_{1}} & 
 m_c A_c & \de_{23}^{ULR} m_{\tilde Q_{2}}m_{\tilde U_{3}}\\
 \de_{{31}}^{ULR}  m_{\tilde Q_{3}}m_{\tilde U_{1}} & 
 \de_{{32}}^{ULR} m_{\tilde Q_{3}} m_{\tilde U_{2}}& m_t A_t 
\end{array}\right)~,
\label{v2Au}
\end{equation}
\noindent 
the matrix ${\cal A}^d$  has a similar form.

We found that the values of all the $\de_{{ij}}^{fAB} $ show a decoupling effect, as it is diplayed in fig.~\ref{fig:DelQLL23} for the case of  $\de_{{32}}^{ULR}$. However, for $\de_{{32}}^{QLL}$ we found a non decoupling effect. The increase of this term with $m_0$ produces important contributions to the  EWPO as we will see in the next section. 

\section{Computation of some observables including squark FV.}
The flavor violating parameters, generated from the RGE runing,  enter at one loop in the computation of the physical observables.  Numerically,  the results
have been obtained using the code 
\fh~\cite{feynhiggs}, which contains the complete set of one-loop corrections from (flavor
violating) squark and slepton contributions as given is ref.~\cite{arana}.
\begin{figure}[ht!]
\begin{center}
  {\includegraphics[scale=0.44]{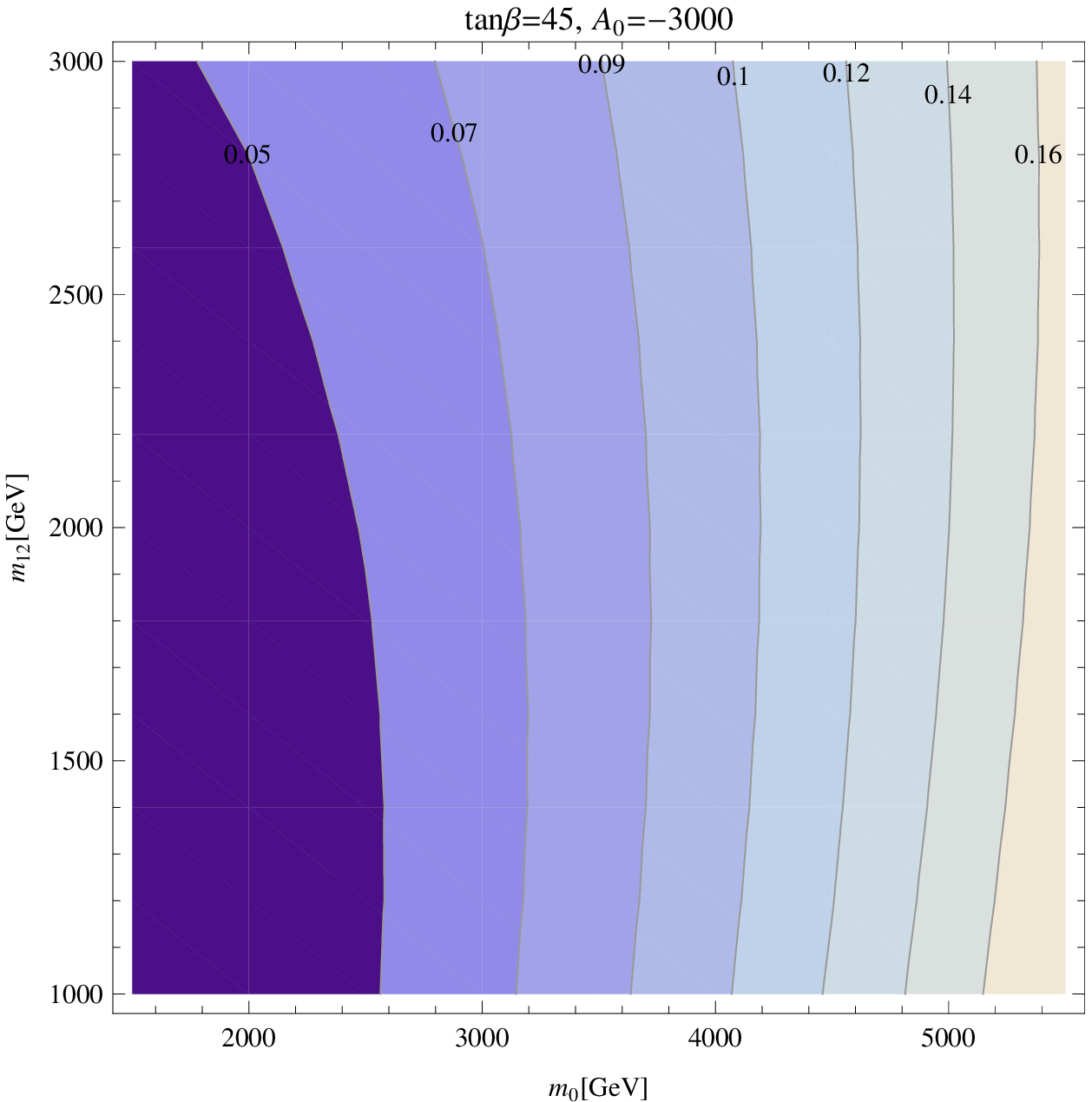}}\hspace{0.5cm}
        {\includegraphics[scale=0.44]{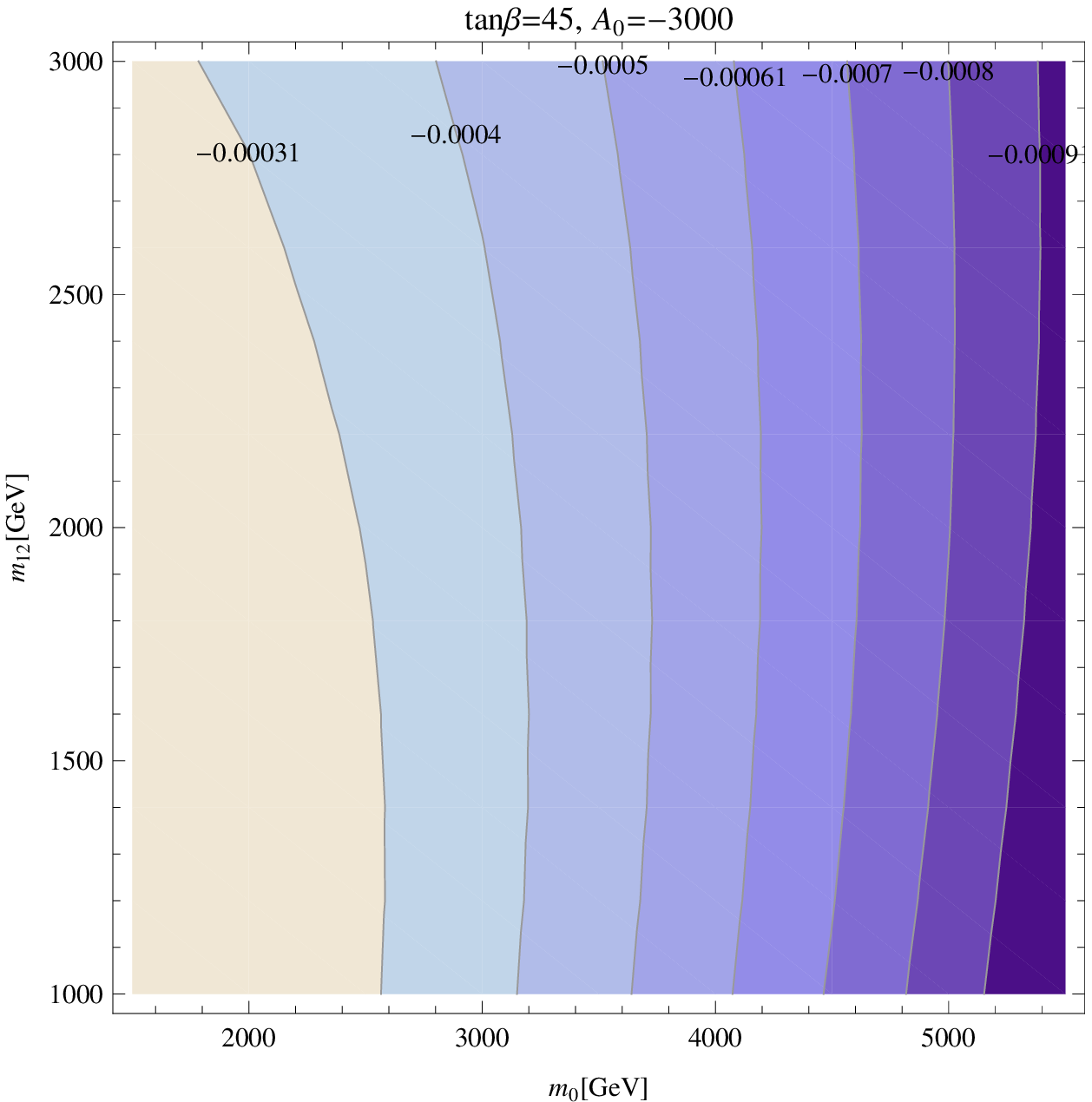}}
\end{center}
\caption{\it Contours of \DMW\ in GeV (left) and  $\sweff$ in the
  $m_0$--$m_{1/2}$ plane for $\tb=45$ and    
$A_0=-3000$~GeV in the CMSSM.}  
\label{fig:SQ-delMW}
\end{figure} 

EWPO, which are known with a great  accuracy, have the potential to allow a discrimination between quantum effects of the SM and SUSY models.  Examples 
are the $W$-boson mass $\MW$ and the $Z$-boson observables, such as the 
effective leptonic weak mixing angle $\sweff$, whose present 
experimental uncertainties are $\de\MW^{\rm{exp,today}} \sim 15 \mev$ and $\de\sweff^{\rm{exp,today}} \sim 15 \times 10^{-5}$.  The experimental uncertanity will further be reduced to $ \sim 4\mev$ and $\sim 1.3 \times 10^{-5}$ respectively in future linear colliders. 

To show explicitely the contribution of the FV entries to the different observables,  we compare the full contribution with the value obtained by setting all $\deFABij = 0$. The results for $\DMW = \MW-\MW^{\rm MSSM}$ and $\Dsweff = \sweff-\sweff^{\rm MSSM}$ (where $\MW^{\rm MSSM}$ and $\sweff^{\rm MSSM}$ are the obtained values with all $\deFABij = 0$) are displayed on fig.~\ref {fig:SQ-delMW}. We can observe a non-decoupling behavoir for the EWPO similar as the one observed in some of the  $\delta^{QLL}_{23}$ as shown in fig. \ref{fig:DelQLL23}. The FV contributions to $\MW$ and $\sweff$ can be above the  experimental uncertainty at some regions of the space of parameters. Therefore, FV contributions can not be neglected in their evaluation. Particularly,   in view of a future improved experimental accuracies.

The FV contribution to other observables turn out to be small, the full FV computation does not lead to significant differences respect the common approach of setting all the $\deFABij = 0$. In the case of the lightest MSSM Higgs boson, the uncertainties arising from the theoretical computation are larger than the exprimental precision of the Higgs mass discovered at the LHC. Even though, we find that the values for 
$\DMH = \MH - \MH^{\rm MSSM}$ that enter in the theoretical prediction  are far below the experimental precision. Similarly, we found that for BPO the approach of taking $\deFABij = 0$ is justified. 

\section{Conclusion}

We studied the impact of including MFV entries on the sfermion mass matrices as they  arise naturally on the CMSSM when the  CKM matrix is included in the RGE's. After a carefull evaluation of several precission observables, we conclude that the effect is not very significant for BPO and Higgs boson masses. However,  EWPO receive contributions that show a non-decoupling behavoir as the values of the SUSY spectrum increases. For instance, those effects can be  larger than the current
experimental accuracy in $\MW$ and $\sweff$. Taking those effects
correctly into account places new upper bounds on $m_0$ that are
neglected in recent phenomenological analyses.  Further applications to FV Higgs decays can be found in ref.~\cite{us3}. Our conclusions can also apply to popular neutrino motivated extensions of the CMSSM like  the \CMSSMI\ .

\section{Acknowledgements}
This work is  supported by by 
the Spanish MICINN's Consolider-Ingenio 2010 Programme under grant
MultiDark CSD2009-00064. The authors also acknowledge further support from  CICYT (S.H. and M.R. from the grant FPA
2013-40715-P and M.E:G. from FPA2011-23781).


\begin{thebibliography}{99}
\bibitem{mssm} 
               H.~Haber and G.~Kane, 
               {\em Phys.\ Rept.} {\bf 117} (1985) 75; 
\bibitem{MFV1} 
 A.~Buras et al., 
 {\em Phys. Lett.} {\bf B 500} (2001) 161.

\bibitem{us1}
  M.~E.~G\'omez, S.~Heinemeyer and M.~Rehman,
  Eur.\ Phys.\ J.\ C {\bf 75} (2015) 9,  434.
 \bibitem{us2}
  M.~E.~G\'omez, T.~Hahn, S.~Heinemeyer and M.~Rehman,
  Phys.\ Rev.\ D {\bf 90} (2014) 7,  074016

\bibitem{Porod:2003um} W.~Porod, 
  {\em Comput. Phys. Commun.} {\bf 153} (2003) 275
  [arXiv:hep-ph/0301101].
\bibitem{feynhiggs} S.~Heinemeyer, W.~Hollik and G.~Weiglein,
                   {\em Comput. Phys. Commun.} {\bf 124} (2000) 76;
                   T.~Hahn, S.~Heinemeyer, W.~Hollik, H.~Rzehak and
                   G.~Weiglein, 
                   {\em Comput.\ Phys.\ Commun.} {\bf 180} (2009) 1426;
  \bibitem{arana} M.~Arana-Catania, S.~Heinemeyer, M.~J.Herrero and S.~Pe\~naranda,
                {\em JHEP} {\bf 1205} (2012) 015
                


\bibitem{us3}
  M.~E.~G\'omez, S.~Heinemeyer and M.~Rehman,
  arXiv:1511.04342 [hep-ph].


\end{thebibliography}
\end{document}